# Using Deep Learning to Predict Neural Stem Cell Differentiation in Regenerative Medicine

**Nidhi Parthasarathy**[*]
Lynbrook High School, San Jose, CA
nidhi.parthasarathy@gmail.com

**Chandra Suda**
Betonville High School, Betonville, AR
kiran1234c@gmail.com

**Anika Mittal**
The Phillips Academy, Andover, MA
anik.mittal@gmail.com

**Ian Young Chen**
The Bronx High School of Science, Bronx, NY
ian9981.chen@gmail.com

**Ananya Jalihal**
Needham High School, Needham, MA
ananyajalihal2006@gmail.com

## Abstract

Over one in three people are affected by neurodegenerative disorders [2]. Neural stem cells, which are multipotent regenerative cells with the potential to differentiate into any of the neural cell types, have immense therapeutic potential for treating neurological disorders. However, lengthy differentiation protocols hinder clinical applications and research. In this study, we present a deep learning approach using convolutional neural networks (CNNs) to predict the fate of neural stem cell differentiation at an early stage. We trained a CNN model on a dataset of cellular images from neural stem cell cultures. Our models achieved impressive results in predicting neuron and glial cell differentiation, with a 93.3% testing accuracy for a multiclass Resnet50 model (and 99.7% accuracy for a binary Resnet50 model). In addition, we developed and published a web tool to give stem cell researchers access to this technology to allow for efficient prediction of stem cell cell differentiation. Our work demonstrates the feasibility of and builds tooling for using CNNs for rapid, early differentiation outcome prediction from simple microscopy images, which could greatly accelerate neural stem cell research and therapies.

## 1 Introduction

Regenerative medicine, or stem cell therapy, has recently received a lot of attention for its tremendous potential [13]. Stem cells – special "building block" cells that can self-renew and differentiate into other types of cells – can be used to repair diseased or damaged tissue, to address numerous diseases (cancers, immuno-deficiencies, genetic diseases), and even grow new tissue to use in organ transplants (especially when organ donors are scarce). *Neural stem cells (NSCs)*, in particular, have been used as a novel treatment strategy for brain tumors, neuro-degenerative disorders, cerebrovascular diseases, strokes, and traumatic brain injury, with broad impact [18]. Alzheimer's disease, for example, the most common cause of dementia, currently impacts 26 million people worldwide, with this number expected to grow to 100 million in the next few decades. Parkinson's is another common neurodegenerative disease that induces motor disorders. Pre-clinical studies using neural stem cells have shown promising results for these and other nervous system disorders [8, 11, 12].

---

[*]All the authors in this paper were interns at MIT's BWSI summer program during the time of this research.

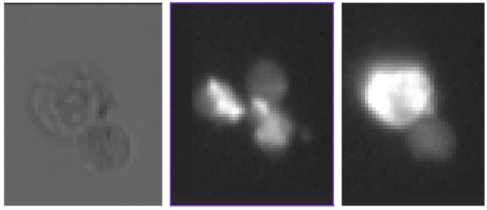

Figure 1: Example images of nerve stem cells used in our dataset

A key aspect of neural stem cells is their ability to differentiate into neurons, astrocytes, and oligodendrocytes, each with different benefits [18]. Differentiating NSCs to neurons can help reconstruct neural circuits damaged by neurological disorders to treat neurodegenerative disease. Differentiating NSCs to astrocytes can help neuroprotection and vascular integrity after injury and help with drug screening. Differentiating to oligodendrocytes can contribute to post-injury re-myelination to help with impulse propagation and metabolic support for neurons.

It is crucial to *track* such differentiation so that therapeutic applications can be efficient about inducing NSCs into specific cell types (usually done through neurotrophins, drugs, hormones, etc). However, common lab methods to observe the effectiveness of inducing differentiation are complex, time consuming, and costly. Deep-tissue methods can be intrusive while other wet-lab methods (immuno-florescent staining, polymerase chain reaction, and western blots) [19] can often takes several days of lab examination and still be error prone depending on molecular marking techniques, lab technology, and operator skill levels [20]. *Automated early prediction of stem cell differentiation* can significantly accelerate experimental iterations and therapeutic development.

AI has emerged as a powerful tool in healthcare with numerous clinical applications of machine learning (neural networks, natural language processing, rule-based systems, robots) [4]. More specific to cell biology, machine learning has been used on images from microscopes or flow cytometry to automatically identify and classify cell types and cell states [5]. For stem cells, one recent study showed that the differentiation process alters the morphology of the stem cells, and such changes can be detected using deep learning models on microscopy data [3]. Another study provides a comprehensive collection of neural stem cells at various stages of differentiation and shows that neural networks can extract spatial features from images to recognize complex patterns and phenotypes [20].

In this study, we study the effectiveness of convolutional neural networks (CNNs) to rapidly predict neural stem cell differentiation based on simple cellular images, to facilitate research and cure of nervous system disorders. CNNs have achieved state-of-the-art performance on computer vision tasks [14] and applying convolutional filters can help extract spatial features from images, enabling the recognition of complex patterns and phenotypes. In addition, we publish our model as a web application that can help the regenerative medicine research community by predicting differentiation and pluripotency potential from cell images.

## 2 Methodology

Our dataset consists of single-cell images from the neural stem cells of embryonic rats captured at various stages of differentiation, available as a public repository from Tongji University (Figure 1). NSCs were immunostained with markers and differentiated into astrocytes, oligodendrocytes, or neurons using specific media with growth factors. Differentiated NSCs were fixed, stained with specific antibodies, and analyzed by flow cytometry, immunofluoresence assays, western blot assays, and RT-transcriptase for cell type identification [20].

The dataset contained 800,000 images, including neural stem cells put through 9 differentiation mediums (astrocyte, oligodendrocyte, retinoic acid/sonic hedgehog, neurotrophin-3, neurotrophin-4, melatonin, nerve growth factor, neurotrophic factor, T3 nanoparticle) and 4 different channels (brightfield, AF488-GFAP, PE-Oligo2, and NeuN-APC). Of these we reduced the dataset to consider the four differentiation mediums with the largest amount of data (astrocyte, oligodendrocyte, retinoic acid/sonic hedgehog, and NT3). This reduced our dataset to around 600,000 samples. To better match our computational resources, we randomly sub-sampled 15% of the data or 90,000 samples. The samples were pre-processed, including image resizing to 45x60, 60x60, or 75x75 depending on the optimal size for the model we were using. For our machine learning models, we used 80% of the data for training and 20% for testing.

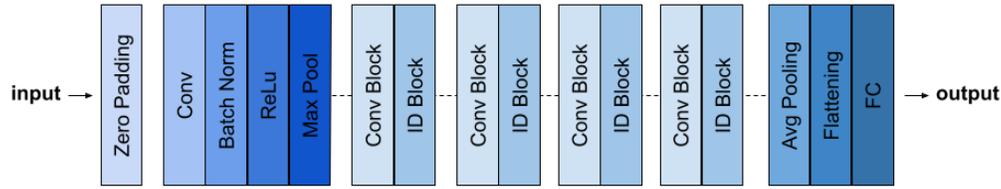

Figure 2: The architecture of the ResNet50 model we studied.

We developed a CNN architecture optimized for classifying NSC differentiation from phase contrast images. The model consists of convolutional layers for feature extraction, pooling layers for spatial reduction, dropouts for regularization, and dense layers for classification. We evaluated different machine learning techniques to classify the differentiation medium of neural stem cells and tried various transfer learning models, specifically focusing on four top models: MobileNet, VGGNet19, InceptionV3, and ResNetv2 (ResNet18 and ResNet50).

*MobileNet* utilizes depthwise separable convolutions to efficiently and accurately compute predictions, and is optimal for lower-sized images [7]. *VGGNet19* is a deep convolutional neural network able to capture intricate features and patterns [15]. *InceptionV3* uses various inception modules designed to capture information at multiple scales and resolutions, well matched with the heterogeneity of image types in our dataset [17]. *ResNetv2* facilitates the training of deeper and more accurate models by creating a "skip connection" across its first and last layers without losing information [6].

Figure 2 shows the architecture of the ResNet model, highlighting the sequence of convolutional (Conv) layers, identity (ID) blocks, and pooling layers, with skip connections in the ID blocks enabling the input to bypass certain layers, thereby facilitating gradient flow and improving training efficiency. We studied both Resnet18 and Resnet50 to evaluate the impact of deeper models. We started with the AveragePooling2D layer which is important in our case because we have 2D grayscale images which are very hard to differentiate with the naked eye, so the high number of filters and complexity of the model allow us to create more useful features. We also used dense and dropout layers to make sure we have a fully connected system and that the model doesn't overfit to the training data.

## 3 Results and Discussion

### 3.1 Metrics

For our models, we studied the overall testing accuracy as the main metric, but also supplemented these with confusion matrices and AUC-ROC curves and scores. Specifically, the testing accuracy measures the number of correctly classified cells among all cells. The *confusion matrix* is a table that compares the predicted labels to the true labels giving insights into the performance of a classification model. The *AUC-ROC* curve above plots the True Positive Rate (sensitivity) against the False Positive Rate (1-specificity) at various threshold settings, providing an accurate representation of a model's performance across all thresholds. The AUC-ROC score captures the total area under the curve and shows the model's capacity to distinguish between classes. A higher AUC-ROC score means the model is better at predicting the classifications correctly.

### 3.2 Results

We first studied a binary classifier focusing on just astrocyte and oligodendrocyte differentiation. MobileNet and ResNet both did well, with MobileNet achieving a 99.7% accuracy rate and a near-perfect AUC-ROC score, and very few false classifications in the confusion matrix.

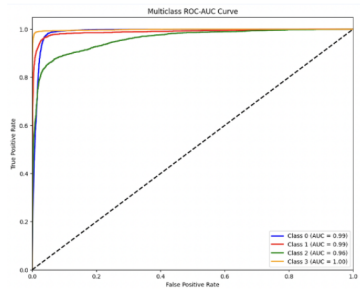

Figure 3: Multiclass ROC-AUC curve for ResNet50 model

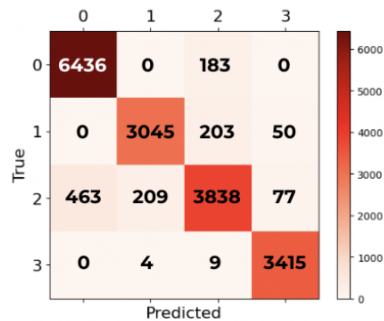

Figure 4: Confusion matrix for ResNet50 model

We studied a multi-class classifier to classify all the four different types of nerve cells in our dataset. Resnet18 and Resnet50 did well. Overall, the Resnet50 model achieved a 93.3% accuracy in classifying undifferentiated NSCs, neurons, and glial cells. Moreover, the Resnet50 model had high AUC-ROC scores for all 4 classes including scores of 0.99, 0.99, 0.96, and 1.00 for NSCs treated with differentiation mediums of astrocyte, oligodendrocyte, retinoic acid/sonic hedgehog, and NT3, respectively (Figure 3). The strong performance demonstrates that deep learning can rapidly predict NSC fate from basic morphology in phase contrast images, compared to traditional approaches involving lengthy immunostaining or functional assays. We also evaluated the performance in a multi-class confusion matrix (Figure 4). The results showed some misclassifications, however the model was highly accurate for most test-samples.

Overall, our results show the ability of deep learning to predict the classification of neural stem cells even in early stages with high accuracy. It is notable that in spite of our subsampled dataset, we still get high accuracies. Our results also show that deeper models can provide improved accuracies.

As an example of how our models can be operationalized for research, we built a web tool at https://stemcells.anvil.app. The user uploads an image which gets processed on the server side with a recommendation from the classifier that is sent back to the client in minutes.

### 3.3 Discussion

With additional computational resources (specifically RAM) and additional pre-processing, our models can be extended to train on larger data sets and more NSC classes (for example, [1, 9]). Hyperparameter tuning for example using Bayesian classifier can also potentially improve accuracies![16]. We are also currently exploring model ensembling (e.g., concatenating across VGGNet and Inception) and large language models (LLAVA [10]).

Our work can be further extended. Convolutional feature visualizations can provide additional insights into defining morphologies of each neural cell type. Additional training data can be added to predict different lineages such as cardiomyocytes, hepatocytes, etc. Live-cell imaging can improve temporal tracking of differentiation. While we focused on differentiating NSC populations, future work can also classify mixed cultures or co-culture systems.

## 4 Conclusion

This work demonstrates deep learning can rapidly predict NSC differentiation fate from simple cellular images. A CNN model achieved exceptional performance in classifying undifferentiated NSCs, neurons, and glial cells (astrocytes and oligodendrocytes). This can significantly accelerate large-scale differentiation experiments by enabling early fate screening. The approach generalizes well to various NSC lines and differentiation protocols. Overall this demonstrates a promising new paradigm for integrating deep learning in stem cell engineering and regenerative medicine. See our website for more detail about our project and our code for the classifiers mentioned above: https://stemcells.anvil.app.

## Acknowledgments and Disclosure of Funding

This work was done as part of an internship at MIT's BWSI Medlytics program. The project was defined and worked on by the student authors of the paper. Christian Cardozo was the instructor for the course and we thank him for his mentoring and feedback in selecting the topic and guidance. We would also like to thank our TAs, Aryan, Catherine, Divya, and Michelle, for their help in clarifying concepts and pointing us to appropriate resources.